# Accurate measurement of $^{13}$C – $^{15}$N distances with solid-state NMR

Jae-Seung Lee and A. K. Khitrin

*Department of Chemistry, Kent State University, Kent, Ohio 44242-0001*

**Abstract**
Solid-state NMR technique for measuring distances between hetero-nuclei in static powder samples is described. It is based on a two-dimensional single-echo scheme enhanced with adiabatic cross-polarization. As an example, the results for intra-molecular distances in α-crystalline form of glycine are presented. The measured NMR distances $^{13}$C(2) – $^{15}$N and $^{13}$C(1) – $^{15}$N are 1.496 ± 0.002 Å and 2.50 ± 0.02 Å, respectively.

## 1. Introduction

Nuclear magnetic resonance (NMR) is a powerful tool for studying molecular structures. Information about inter-nuclear distances is retrieved from magnitudes of long-range dipole-dipole interactions. In liquids, where dipolar couplings are averaged out by fast molecular motions, inter-atomic distances can be estimated from nuclear Overhauser effect.[1] More quantitative information is obtained with using anisotropic liquid-crystalline solvents,[2] which partially restore the dipole-dipole couplings. Orientation with bicelles became an efficient technique to determine structures of complex biomolecules.[3] For molecules of liquid crystal itself, dipolar couplings provide valuable information about structure and ordering.[4-6]

More accurate data on inter-nuclear distances can be obtained for solids, where the effect of molecular motions is small. Various techniques, like separated-local-field spectroscopy[7] or PISEMA,[8,9] have been developed for measuring distances between rare and abundant nuclei. Since quality of heteronuclear decoupling can be substantially higher than that of homonuclear decoupling, one would expect better accuracy for well-isolated pairs of rare nuclei when abundant nuclei (protons) are decoupled. Today, most of the solid-state NMR experiments use magic-angle spinning (MAS) of a sample, which, by eliminating chemical shift anisotropy in powder samples, solves two problems: allows site assignments and increases signal-to-noise (S/N) ratio by creating narrow spectral peaks. MAS also averages the dipole-dipole couplings, and they should be reintroduced by using a rotational resonance.[10] Rotor-synchronized recoupling sequences[11] make it possible to combine sites assignment with information about dipolar couplings. When the pairs of rare nuclei are created by specific isotope labeling and, therefore, the problem of assignment is solved, the most popular techniques for measuring inter-nuclear distances under MAS are REDOR[12] (for heteronuclear pairs) and build-up of the double-quantum coherence[13] (for homonuclear pairs). In both of these techniques the experimental spin dynamics under MAS is compared to results of computer simulation. Such simulation requires additional information about tensors of chemical shifts and their orientations with respect to the internuclear vector.

Typical accuracy of measuring the constants of dipole-dipole interaction under MAS is about 10%. More accurate NMR measurements have been performed for static samples. Among the techniques are the nutation spectroscopy[14] or use of the Carr-Purcell sequence[15] for eliminating the chemical shift interaction in homonuclear systems, and MLEV-8 sequence with composite pulses[16] in heteronuclear systems. When continuous or multi-pulse irradiation is strong (but, in practice, not infinitely strong) the system evolves under the influence of an effective Hamiltonian, which is known only approximately. To solve this problem, in[15] the experimental data were extrapolated to infinite intervals between pulses in a Carr-Purcell sequence. However, there is a problem with such extrapolation for homonuclear systems: the frequency of dipolar oscillations of a two-spin system depends on a difference of chemical shifts. For the same orientation of the internuclear vector, say perpendicular to the magnetic field, different orientations of crystals, allowed by rotations around this internuclear vector, create a distribution of the chemical shift differences, if the chemical shift tensors are not exactly the same for both nuclei. The situation is much better for heteronuclear pairs when the dipole-dipole interaction is truncated to ZZ-term and commutes with the difference of chemical shifts. Therefore, the extrapolation to infinite intervals between pulses for a heteronuclear system[16] is conceptually correct. At the same time, in both cases of using the extrapolation to infinite intervals between pulses[15,16] the difference between actually measured and extrapolated values exceeded 10%.

We believe that better accuracy can be achieved when perturbation of the system is physically reduced to minimum and most of the time the system evolves under the unperturbed Hamiltonian of a heteronuclear two-spin system

$$H = \omega_I I_z + \omega_S S_z + \left[\frac{\mu_0}{4\pi}\frac{\gamma_I \gamma_S \hbar}{r^3}\left(1 - 3\cos^2\theta\right) + J\right] I_z S_z, \qquad (1)$$

where $\omega_I$ ($\omega_S$) and $\gamma_I$ ($\gamma_S$) are the chemical shift and the gyromagnetic ratio of spin $I$ ($S$), $\mu_0$ is the permeability of free space, $r$ is the distance between spins $I$ and $S$, $\theta$ is the angle between the static magnetic field and the internuclear vector connecting spins $I$ and $S$, and $J$ is the isotropic spin-spin coupling constant. Refocusing of the chemical shift interactions can be performed with a single pair of simultaneous 180° pulses on both of the heteronuclei. Insufficient sensitivity may be the major difficulty in detecting static powder Pake doublets from spin pairs at low concentration. However, as we demonstrate below, a combination of adiabatic cross-polarization[17] and efficient heteronuclear decoupling[18-21] makes it possible to study even strongly-coupled $^{13}C - ^{15}N$ pairs at concentrations 1% or less.

## 2. Scheme of experiment

The scheme of experiment is shown in Fig. 1. It starts with adiabatic cross-polarization[17] (ACP) to boost polarization of the $^{13}C$ nuclei. ACP consists of adiabatic demagnetization/remagnetization in the laboratory frame (ADLF/ARLF), performed with two frequency-sweeping pulses. The efficiency of ACP is illustrated in Fig. 2, showing about seven-fold increase of the $^{13}C$ polarization in glycine.

After ACP, $^1H$ spins are decoupled, and a 90° pulse creates transverse magnetization of $^{13}C$ spins. The spin system consisting of $^{13}C$ and $^{15}N$ spin pairs evolves under their dipole-dipole interaction, while the two simultaneous 180° pulses on both spins in the middle of the evolution period $t_2$ refocus dephasing by chemical shifts and form an echo at the end of the evolution period. The $^{13}$C NMR signal is acquired in a two-dimensional way by varying the evolution time. The Fourier-transform with respect to the evolution time $t_2$ gives the $^{13}$C – $^{15}$N dipolar powder spectrum.

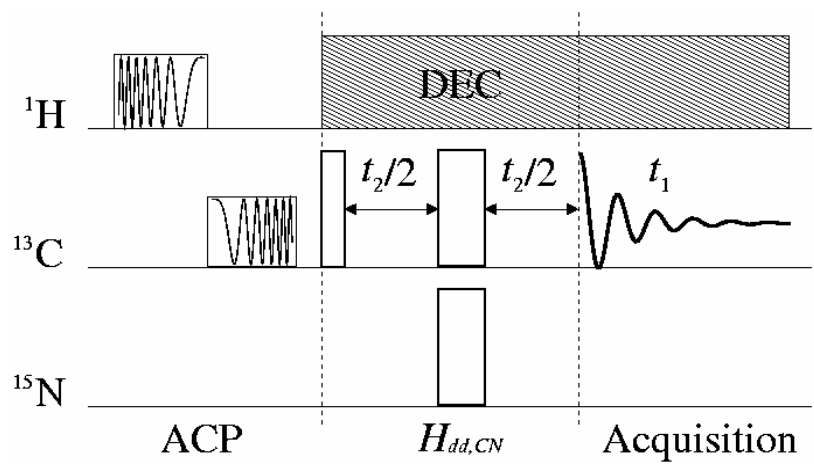

Fig. 1 Experimental pulse sequence consists of adiabatic cross-polarization, evolution ($t_2$), and signal acquisition ($t_1$) periods.

## 3. Samples and experimental parameters

To test this scheme we have used α-crystalline form of glycine which has been previously studied with various techniques including NMR,[22,23] X-ray[24] and neutron scattering.[25-27] Isotopically labeled glycine-[2-$^{13}$C,$^{15}$N] (ISOTEC) has been used as a test for short $^{13}$C – $^{15}$N distance, and glycine-[1-$^{13}$C,$^{15}$N] (Aldrich) for longer $^{13}$C – $^{15}$N distance. Labeled glycine was diluted to 2% in natural abundance (n. a.) glycine (Aldrich) and to 1% in isotopically depleted glycine-[$^{12}$C$_2$,$^{14}$N] (ISOTEC). All four samples were recrystallized from aqueous solutions. The experiments have been performed at room temperature (25°C) with a Varian Unity/Inova 500 MHz NMR spectrometer using a triple resonance "indirect" probe for liquids. In the ACP period, the first $^1$H adiabatic pulse had 100 ms duration, 100 kHz frequency-sweeping range and 1.2 kHz amplitude ($\gamma B_1/2\pi$); the second $^{13}$C pulse had 100 ms duration, 60 kHz frequency-sweeping range and 9.5 kHz amplitude. In the evolution period, the 90° and 180° pulses on $^{13}$C spin were, respectively, 16 μs and 32 μs long, and the 180° pulse on $^{15}$N spin was 80 μs long. The number of $t_2$ increments was 128, and the longest echo time was 50.8 ms. Proton spins were decoupled during the evolution and acquisition periods by using the SPINAL-32 heteronuclear decoupling sequence.[19] The signal acquisition time $t_1$ was set to 2 ms.

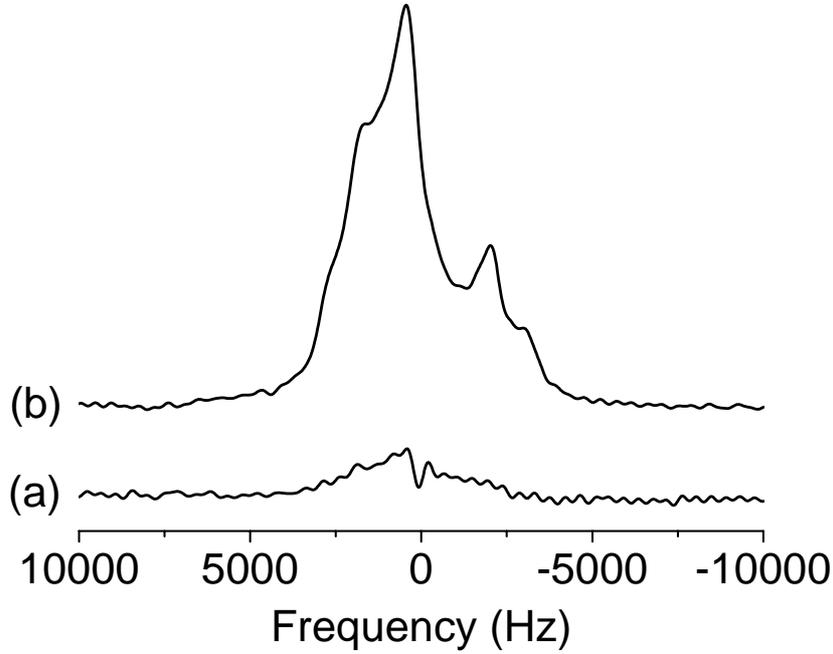

Fig. 2. $^{13}$C NMR spectrum of 2% glycine-[2-$^{13}$C,$^{15}$N] in n. a. glycine: (a) thermal equilibrium spectrum; (b) the spectrum enhanced with adiabatic cross-polarization.

## 4. Results

Fig. 3 shows the dipolar powder spectra obtained for (a) 1% glycine-[2-$^{13}$C,$^{15}$N] in isotopically depleted glycine-[$^{12}$C$_2$,$^{14}$N], (a') 2% glycine-[2-$^{13}$C,$^{15}$N] in natural-abundance glycine, (b) 1% glycine-[1-$^{13}$C,$^{15}$N] in glycine-[$^{12}$C$_2$,$^{14}$N], and (b') 2% glycine-[1-$^{13}$C,$^{15}$N] in natural-abundance glycine. Both the "singular" peaks contributed by the internuclear vectors perpendicular to the magnetic field $\left(\theta \approx \frac{\pi}{2}\right)$ and step-like boundaries, which come from the parallel orientation $(\theta \approx 0)$, are clearly seen in the dipolar powder spectra. The dipolar frequency $\omega_D = \frac{\mu_0}{4\pi}\frac{\gamma_I \gamma_S \hbar}{r^3}$ and, therefore, NMR distance $r_{NMR} = \left(\frac{\mu_0}{4\pi}\frac{\gamma_I \gamma_S \hbar}{\omega_D}\right)^{1/3}$ can be extracted from the spectra in Fig. 3 by using a simple data processing described in the following section.

## 5. Data processing

There are three major factors that create distortions of the spectra in Fig. 3 and make them differ from the ideal theoretical dipolar powder spectra: line-broadening, which mostly comes from non-perfect proton decoupling in our experiments; scalar *J*-coupling, which should be taken into account for the case of the nearest neighbors (Figs. 3a and 3a'); and the delayed acquisition in $t_2$ dimension, which is due to finite pulses duration. Intensity of the measured signal decays with increasing evolution time $t_2$. This line-broadening shifts the "singular" peaks at $\pm\omega_D/2$ toward the center of the spectrum. Under the assumption that the signal decays exponentially with $t_2$, the spectrum is a convolution of the ideal dipolar powder spectrum and the Lorentzian shape. Simulated theoretical spectra for three different broadening factors are shown in Fig. 4. Even though the peak positions shift toward the center of the spectrum at increasing broadening factor, one can notice that all the spectra cross the frequency $\pm\omega_D/2$ at the same relative height, about 90% of the maximum (see the inset in Fig. 4). This suggests a simple method of correcting for the broadening factor: the spectral interval should be measured between the two outside points at 90% of relative intensity.

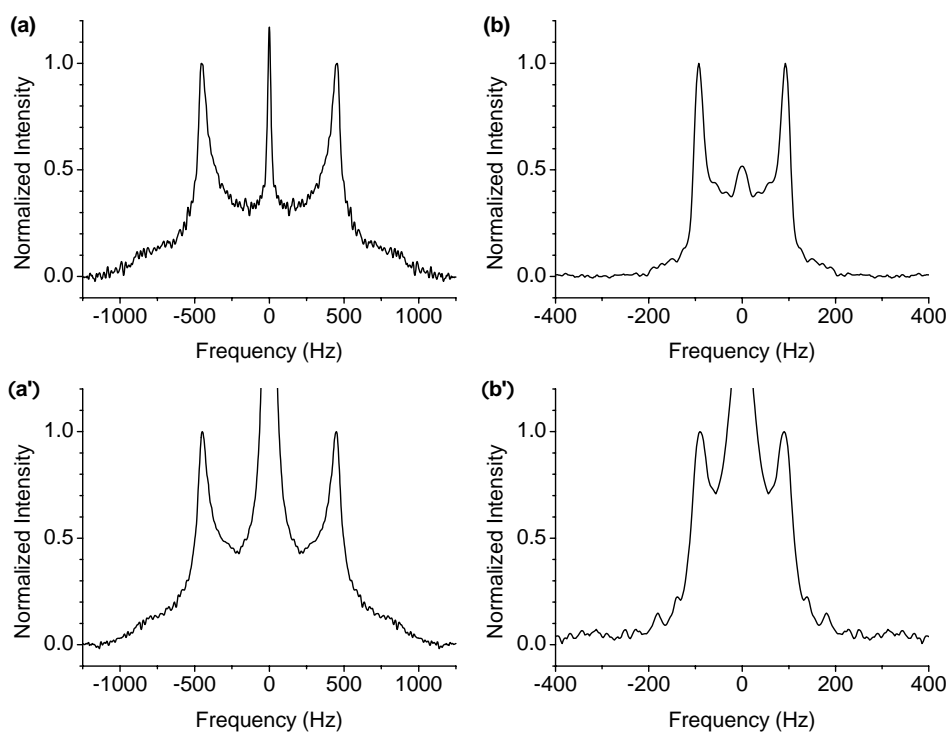

Fig. 3 Dipolar spectra for (a) 1% glycine-[2-$^{13}$C,$^{15}$N] in glycine-[$^{12}$C,$^{14}$N] (exp. time ~ 16.5 hours); (a') 2% glycine-[2-$^{13}$C,$^{15}$N] in n. a. glycine (exp. time ~ 20.5 hours); (b) 1% glycine-[1-$^{13}$C,$^{15}$N] in glycine-[$^{12}$C,$^{14}$N] (exp. time ~ 20.5 hours); (b') 2% glycine-[1-$^{13}$C,$^{15}$N] in n. a. glycine (exp. time ~ 20.5 hours).

For the directly bonded $^{13}$C and $^{15}$N nuclei, the scalar *J*-coupling should also be taken into account. Although the value of *J*-coupling can be obtained from fitting the spectrum in Fig. 3a, we preferred not to introduce an extra fitting parameter and used the value $J = 6.7$ Hz measured separately in a solution. The signs of *J*-coupling and the dipolar coupling at $\theta = \pi/2$ are the same (negative). Therefore, 6.7 Hz was subtracted from the frequency interval between the points at 90% height.

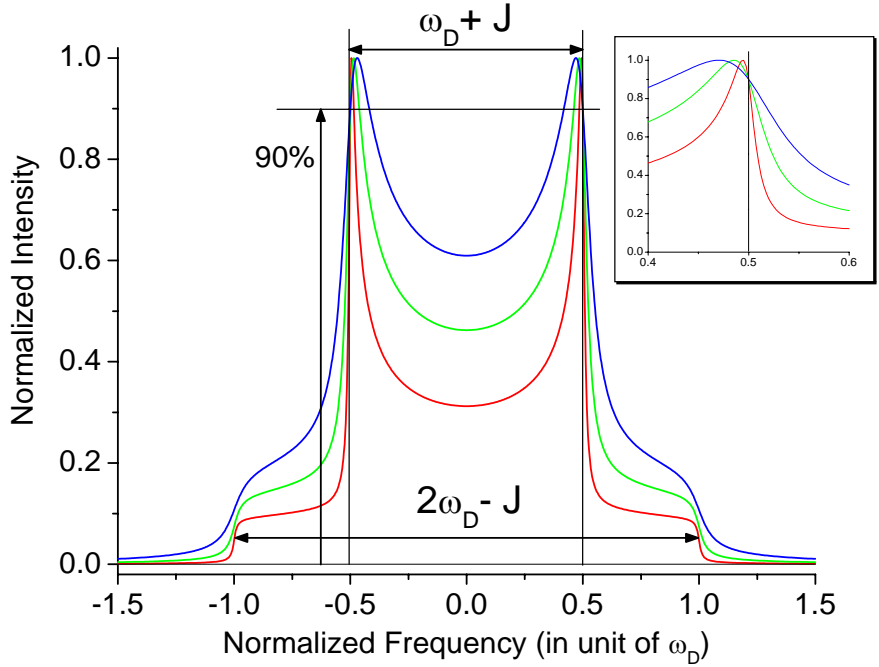

Fig. 4 Simulated spectra for different relative broadening factors: 2% (red), 5% (green), and 10% (blue). Although the peak positions shift with increasing broadening factor, the spectral intervals between the points at 90% of relative peak intensity can be consistently measured, as shown in the inset.

The distortion from the pair of 180° pulses is small but not zero. Finite duration of the pulses and after-pulse delays create an acquisition delay $\tau$ in the $t_2$ dimension. In the absence of line broadening, the acquisition delay affects the shape of the spectrum but does not shift the positions of the singular peaks. Our numerical study showed that there exists interference between line-broadening and acquisition delay, which shifts the peaks outside. The corresponding frequency shift is well described by the following empirical formula:

$$\frac{\Delta\omega}{2\pi} = 2.2 \times \frac{\omega_D}{2\pi} \times (4 \times \Delta f) \times \tau , \qquad (2)$$

where $\Delta f$ is the distance in frequency between the points of 90% and 70% intensities on the outside slope of a "singular" peak. The Lorentzian broadening factor is $4\Delta f$.

The step-by-step results following this data processing scheme are shown in Table 1. First, the distances between the points of 90% intensity are determined and adjusted by subtracting the spin-spin coupling constant $J$. This value is denoted $\omega'_D$ in the Table 1. Then, the line broadening factors are estimated from the frequency interval between the points of 90% and 70% intensities on the outside slope of a "singular" peak. The additional shift, resulting from the acquisition delay, is calculated with Eq.(2) and subtracted from $\omega'_D$ to give $\omega_D = \omega'_D - \Delta\omega$. The internuclear distances $r_{NMR}$ are calculated as $\left( \dfrac{\mu_0}{4\pi} \dfrac{\gamma_I \gamma_S \hbar}{\omega_D} \right)^{1/3}$.

Table 1. Experimental results for intra-molecular $^{13}C - ^{15}N$ distances in α-glycine. $(\omega'_D+J)/2\pi$ is the frequency interval at 90% height (see text), $4\times\Delta f$ is the measured line-broadening factor, and $\omega_D/2\pi$ is the dipolar coupling constant after correction for the acquisition delay. The data in parentheses are obtained with $^{15}N$ detection.

| Sample | $(\omega'_D+J)/2\pi$ (Hz) | $4\times\Delta f$ (Hz) | $\omega_D/2\pi$ (Hz) | $r_{NMR}$ (Å) |
|---|---|---|---|---|
| 1% glycine-[2-$^{13}$C,$^{15}$N] in glycine-[$^{12}$C,$^{14}$N] | 929.21 (924.64) | 35.9 (14.8) | 915.21 (914.29) | 1.4959 (1.4964) |
| 2% glycine-[2-$^{13}$C,$^{15}$N] in n. a. glycine | 925.67 (923.40) | 59.4 (17.9) | 906.97 (912.28) | 1.5004 (1.4975) |
| 1% glycine-[1-$^{13}$C,$^{15}$N] in glycine-[$^{12}$C,$^{14}$N] | 195.62 (196.07) | 15.9 (13.4) | 194.94 (195.36) | 2.5048 (2.5029) |
| 2% glycine-[1-$^{13}$C,$^{15}$N] in n. a. glycine | 197.80 (196.22) | 25.8 (15.5) | 196.68 (195.40) | 2.4973 (2.5028) |

A comparison of the spectrum in Fig. 3a with the theoretical one is presented in Fig. 5. In the theoretical spectrum, we used the experimental value of the acquisition delay τ = 100 μs, which is a sum of the $^{15}$N pulse duration (80 μs) and two after-pulse delays (10 μs each). Some deviations between the theoretical and experimental spectra are due to the fact that the line-broadening factor depends on crystals orientation and, as a result, the total spectrum cannot be fitted using a uniform line broadening. This is one of the reasons why we used narrow spectral regions of the outside slopes of the "singular" peaks for retrieving information about dipolar coupling constants, rather than trying to extract this information from fitting the total powder pattern.

Fitting with variable $\omega_D$ and $\Delta f$ of the outside slopes of the "singular" peaks of the spectrum in Fig. 3a was used to estimate a statistical error of the results. It showed very small uncertainties of ± 1 Hz for $\omega_D$ and ± 3 Hz for $\Delta f$, determined as deviations from the best fit values where the r.m.s. difference between the experimental and theoretical spectra doubles.

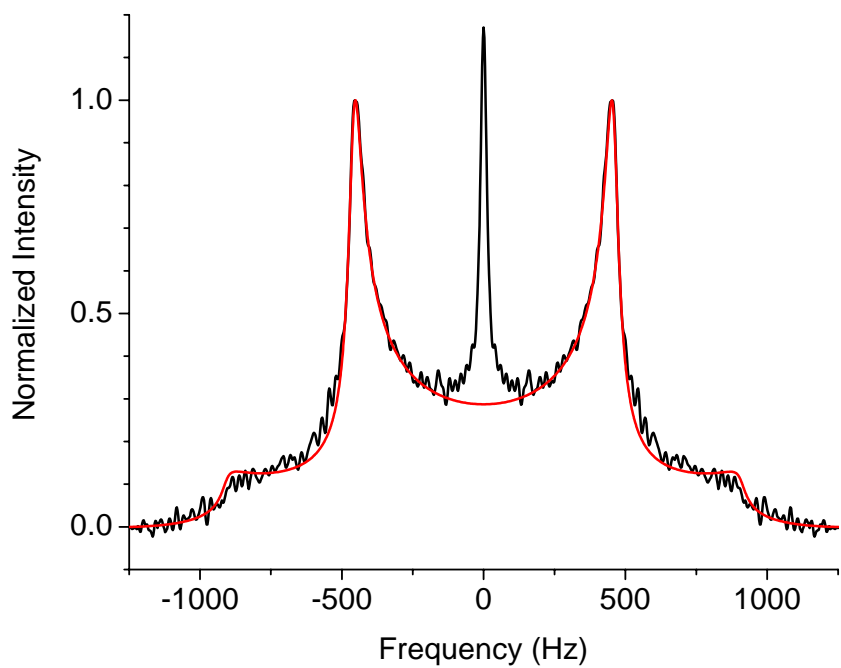

Fig. 5 Comparison of experimental (black) and theoretical (red) powder spectra for 1% glycine-[2-$^{13}$C,$^{15}$N] in isotopically depleted glycine-[$^{12}$C,$^{14}$N].

Correlation between chemical shifts and orientations of the internuclear vector, in the presence of line broadening, create small differences between traces of the two-dimensional spectrum. The spectra shown in Fig.3 are so-called "sky projections" obtained with the Varian's VNMRJ software. The values of $\omega_D$ calculated from different traces are distributed in a range ± 5 Hz. Since we did not analyze the correlation between chemical shifts and orientations of the internuclear vector, we treated this difference between traces as a source of error and estimated, conservatively, the relative errors in $r_{NMR}$ to be about 0.15 % and 0.8 % for the short and long distances, respectively.

## 6. Atomic vibrations

The dipolar coupling constant $\omega_D$ measured by NMR is a quantity averaged over molecular motions. Modulation of in the distance $r$ between nuclei and the angle $\theta$ affect the measured $\omega_D$ and, therefore, the calculated distance $r_{NMR}$. This effect can be clearly seen when one compares the distances measured by NMR with the ones obtained from X-ray or neutron scattering (Table 2). When $r_{NMR}$ is determined from positions of the "singular" peaks in the powder spectrum $\left(\theta \approx \dfrac{\pi}{2}\right)$ one can introduce the deviations $r = r_0 + \Delta$, $|\Delta| \ll r_0$, and $\theta \equiv \dfrac{\pi}{2} - \delta$, $|\delta| \ll 1$, and estimate the effect of atomic vibrations on $r_{NMR}$ as

$$r_{NMR} \equiv \left(\frac{\gamma_I \gamma_S \hbar}{\omega_D}\right)^{1/3} = r_0 \left(1 + 6\frac{\langle (\Delta)^2 \rangle}{r_0^2} - 3\langle \delta^2 \rangle\right)^{-1/3} \approx r_0 \left(1 - 2\frac{\langle \Delta^2 \rangle}{r_0^2} + \langle \delta^2 \rangle\right), \quad (3)$$

where $r_0$ is the equilibrium distance between nuclei. Alternatively, the equilibrium distance $r_0$ can be calculated from $r_{NMR}$:

$$r_0 \cong r_{NMR}\left(1 - \langle \delta^2 \rangle\right) + \frac{2\langle \Delta^2 \rangle}{r_{NMR}} \quad (4)$$

when the variances of $\Delta$ and $\delta$ are known. C – N bond is rigid, and it is reasonable to assume that C – N stretching is close to one of the crystal normal modes. Then, $\langle \Delta^2 \rangle$ can be estimated from temperature and frequency of the corresponding harmonic oscillator. For an oscillator, $\dfrac{1}{2}m\omega^2 \langle \Delta^2 \rangle = \dfrac{1}{2}E_{tot}$, where $m$ is the reduced mass, $\omega$ is the frequency, and $E_{tot}$ is the total energy. The average energy of harmonic oscillator is $\overline{E} = \hbar\omega\left[\dfrac{1}{2} + \dfrac{1}{\exp(\hbar\omega/k_B T) - 1}\right]$ and, therefore,

$$\langle \Delta^2 \rangle = \frac{\overline{E}}{m\omega^2} = \frac{\hbar}{m\omega}\left[\frac{1}{2} + \frac{1}{\exp(\hbar\omega/k_B T) - 1}\right]. \quad (5)$$

At $\hbar\omega \gg k_B T$ $\langle \Delta^2 \rangle \cong \dfrac{\hbar}{m\omega}\left[\dfrac{1}{2} + \exp\left(-\dfrac{\hbar\omega}{k_B T}\right)\right]$, and at $\hbar\omega \ll k_B T$ $\langle \Delta^2 \rangle \cong \dfrac{k_B T}{m\omega^2}$. From the IR spectrum of glycine, it is known[27] that the line at 1032 cm$^{-1}$ represents the C-N stretching. Using Eq.(5) one can find that C-N stretching gives $\langle \Delta^2 \rangle = 2.378 \times 10^{-3}$ Å$^2$. This is a small quantity and produces only 0.3% decrease of $r_{NMR}$ compared to $r_0$. Since $\hbar\omega > k_B T$ for this vibration, the increase is mostly contributed by a quantum uncertainty in $r$. It is more difficult to estimate $\langle\delta^2\rangle$, which increases $r_{NMR}$ compared to $r_0$, because it may have contributions from many low-frequency thermally-excited modes. According to molecular dynamics simulation,[28] the average variance of the Euler angles for molecular librations of the C – N bond in glycine is $\langle\delta^2\rangle = 0.01273$ rad$^2$. It would make $r_{NMR}$ longer than $r_0$ by about 1% or 0.02 Å, which is comparable to the differences between the results obtained with NMR and other techniques (Table 2).

Table 2. Comparison of experimental results for C(2) – N distance in α-glycine.

| | | |
|---|---|---|
| Solid State NMR | This work | 1.496 ± 0.002 Å |
| | NMR, $^{13}$C-$^{15}$N, ref.[23,28] 10% glycine-[2-$^{13}$C,$^{15}$N] in n. a. glycine | 1.505 (-0.001, 0.002) Å |
| | NMR, $^{13}$C-$^{14}$N, single crystal, ref.[22] 9% glycine-[1,2-$^{13}$C$_2$] in n. a. glycine | 1.509 ± 0.009 Å |
| X-ray & Neutron | Neutron Scattering I, ref.[25] | 1.476 ± 0.001 Å |
| | Neutron Scattering II, ref.[26] | 1.475 ± 0.001 Å |
| | X-ray Diffraction, ref.[24] | 1.474 ± 0.005 Å |

**7. $^{15}$N detection**

With further improved sensitivity it may become advantageous to use $^{15}$N detection. Lower γ of the $^{15}$N nucleus results in decreased broadening factors, and lower natural abundance makes the results obtained in natural-abundance matrices almost as good as in isotopically depleted matrices. The scheme of experiment is the same as in Fig.1 except that $^{13}$C is replaced by $^{15}$N and vice versa. The adiabatic $^{15}$N pulse had 30 ms duration, 30 kHz sweeping range, and 1.3 kHz amplitude. The dipolar powder spectra for the four glycine samples discussed above are presented in Fig. 6. Because of higher noise level the powder patterns are not distinguishable for the spectra in Figs. 6a and 6a'. At the same time, significantly reduced line broadening makes the "singular" peaks clearly visible well above the noise level. We believe that the data obtained with $^{15}$N detection and shown in parentheses in Table 1 provide more accurate results than $^{13}$C detection. However, we did not attempt to use the $^{15}$N data for reducing the error margins of the presented $r_{NMR}$ values because of high noise level of the spectra in Fig. 6.

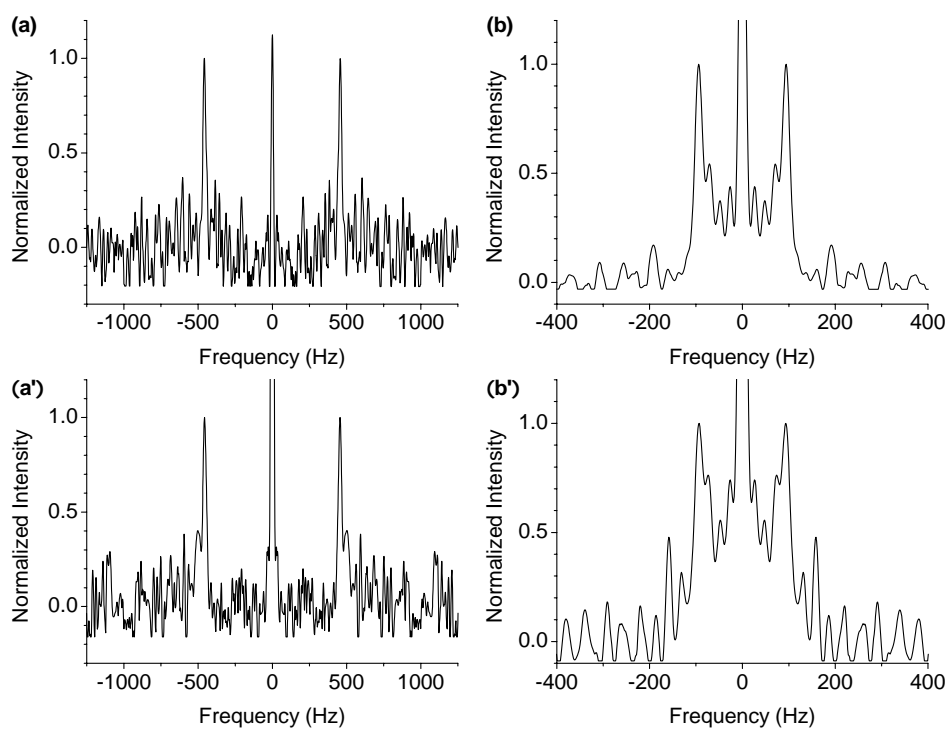

Fig. 6 Dipolar powder spectra obtained with $^{15}$N detection: (a) 1% glycine-[2-$^{13}$C,$^{15}$N] in glycine-[$^{12}$C,$^{14}$N] (exp. time ~ 80 hours); (a') 2% glycine-[2-$^{13}$C,$^{15}$N] in n. a. glycine (exp. time ~ 47 hours); (b) 1% glycine-[1-$^{13}$C,$^{15}$N] in glycine-[$^{12}$C,$^{14}$N] (exp. time ~ 90 hours); (b') 2% glycine-[1-$^{13}$C,$^{15}$N] in n. a. glycine (exp. time ~ 40 hours).

## 8. Conclusion

The major factor limiting an accuracy of NMR distance measurements in our experiments was line broadening, resulting from non-perfect protons decoupling. We used a conventional probe for liquids and the available protons decoupling power was only 50 kHz. We think that the accuracy can be considerably increased with using a customized probe, which will have optimized sensitivity for a directly detected nucleus and, at the same time, will allow high-power protons decoupling. In fact, the accuracy of the NMR distance measurement can be so high that such data can be used for refining structures obtained with X-ray or neutron scattering. However, molecular vibrations affect differently the results of different methods, and more work on quantifying the vibrational contributions is needed before different techniques can complement each other.

**Acknowledgements**
The authors thank Profs. C. Jaroniec and M. Tubergen for helpful discussions and Dr. M. Gangoda for technical assistance. The work was supported by Kent State University and Ohio Board of Regents.